\documentclass[fleqn,usenatbib]{mnras}

\usepackage{newtxtext,newtxmath}
\usepackage[T1]{fontenc}

\DeclareRobustCommand{\VAN}[3]{#2}
\let\VANthebibliography\thebibliography
\def\thebibliography{\DeclareRobustCommand{\VAN}[3]{##3}\VANthebibliography}

\usepackage{graphicx}	%
\usepackage{amsmath}	%
\usepackage{cleveref}
\usepackage{physics}
\usepackage[separate-uncertainty=true]{siunitx}
\usepackage{booktabs}
\usepackage{threeparttable}
\usepackage{color}
\usepackage{xcolor}
\usepackage{xspace}
\renewcommand{\vec}[1]{\vb*{#1}}
\crefname{figure}{Fig.}{Figs}
\Crefname{figure}{Fig.}{Figs}
\crefname{table}{Table}{Tables}
\Crefname{table}{Table}{Tables}
\crefname{equation}{Eq.}{Eqs}
\Crefname{equation}{Eq.}{Eqs}

\AtBeginDocument{\RenewCommandCopy\qty\SI}

\DeclareSIUnit\Msun{\ensuremath{\text{M}_\odot}}
\DeclareSIUnit\hred{\textit{h}}

\newcommand{\hagn}{{\sc Horizon-AGN}\xspace}

\newcommand{\orcid}[1]{\href{https://orcid.org/#1}{\includegraphics[height=.7em]{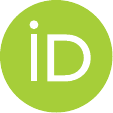}}}

\title[Dimensionality of wide-band photometric data]{How complex are galaxies? A non-parametric estimation of the intrinsic dimensionality of wide-band photometric data}
\author[C. Cadiou et al]{
\orcid{0000-0003-2285-0332} Corentin Cadiou,$^{1,2}$\thanks{E-mail: corentin.cadiou@iap.fr}
\orcid{0009-0008-5926-818X} Clotilde Laigle,$^{2}$
and
\orcid{0000-0002-4287-1088} Oscar Agertz$^{1}$
\vspace{.5em}
\\
$^{1}$Lund Observatory, Division of Astrophysics, Department of Physics, Lund University, Box 43, SE-221 00 Lund, Sweden\\
$^{2}$Sorbonne Université, CNRS, UMR 7095, Institut d'Astrophysique de Paris, 98 bis Boulevard Arago, F-75014 Paris, France\\
}

\date{Accepted 2025 January 22. Received 2024 December 3; in original form 2024 April 4}

\pubyear{2024}

\begin{document}
\label{firstpage}
\pagerange{\pageref{firstpage}--\pageref{lastpage}}
\maketitle

\begin{abstract}
    Galaxies are complex objects, yet the number of independent parameters to describe them remains unknown.
    We present here a non-parametric method to estimate the intrinsic dimensionality of large datasets.
    We apply it to wide-band photometric data drawn from the COSMOS2020 catalogue and a comparable mock catalogue from the \hagn{} simulation. {Our galaxy catalogues are limited in signal-to-noise ratio in all optical and NIR bands.}
    Our results reveal that most of the variance in the wide-band photometry of this galaxy sample can be described with at most \num{4.3(0.5)} independent parameters for star-forming galaxies and \num{2.9(0.2)} for passive ones, both in the observed and simulated catalogues.
    We identify one of these parameters to be noise-driven, and recover that stellar mass and redshift are two key independent parameters driving the magnitudes.
    Our findings support the idea that wide-band photometry does not provide more than one additional independent parameter for star-forming galaxies. Although our sample is not mass-limited and may miss some passive galaxies due to our cut in SNR,
    our work suggests that dimensionality reduction techniques may be effectively used to explore and analyse wide-band photometric data, provided the used latent space is at least four-dimensional.
\end{abstract}


\section{Introduction}
\label{sec:introduction}
In the past decades, we have collected exquisite multi-wavelength photometric data for millions of galaxies in the observable Universe, {e.g. on COSMOS \citep{scoville07} and SXDS \citep{mehta18} fields, or through the HSC-SSP \citep{Aihara18} and DES \citep{hartley22} surveys, among others}. The pace of observations is now accelerating and current and next generations of space-based instruments, {including JWST \citep{gardner06}, \emph{Euclid} \citep{euclid11}, {the Vera Rubin} \citep{lsst19} and {the Nancy Grace Roman} \citep{spergel15} observatories}, will provide in the next few years broad-band optical and near-infrared fluxes for almost all galaxies in the observable Universe over at least the last 10 billion years of history. These photometric datasets represent a goldmine to constrain galaxy formation, evolution and quenching from their photometry.

Galaxy photometric fluxes are directly sensitive to the complex age and metallicity distributions of the stellar population that comprise them. %
Traditionally, astronomers have focused on separating star-forming and passive populations based on two-dimensional colour-colour and colour-magnitude diagrams \citep[CMDs; e.g.][]{deVaucouleurs1961,stratevaColorSeparationGalaxy2001,bellOpticalNearInfraredProperties2003,bell04,arnoutsEncodingInfraredExcess2013}. Recently,  more sophisticated techniques have been explored to efficiently reduce dimensionality while preserving the topology of data, trying to unfold as much as possible the high-dimensional space of galaxy colours into a single two-dimensional manifold.

For example, self-organising maps \citep[SOMs,][]{kohonenSelforganizedFormationTopologically1982,kohonenSelfOrganizingMaps2001} are now widely used as a way to circumvent SED-fitting \citep[see e.g.][]{hemmatiBringingManifoldLearning2019}. In this approach, a SOM is first built as an unsupervised classification of galaxies based on their photometric colours. In a second step, provided a representative fraction of galaxies have a label of the quantity that we want to predict, e.g. redshift, mass, star formation rate (SFR), etc., this information can be efficiently propagated to the rest of the sample. This technique was tested in simulations \citep[e.g.][]{davidzonHORIZONAGNVirtualObservatory2019} and applied to observed datasets to predict redshift \citep{Wilson2020}, SFR \citep{davidzonCOSMOS2020ManifoldLearning2022}, or redshift distribution in tomographic bins for weak lensing studies, such as DES and \textit{Euclid} \citep[see e.g.][]{buchsPhenotypicRedshiftsSelforganizing2019,mylesDarkEnergySurvey2021}.
As an efficient way to compare the multi-wavelength colour distribution of two galaxy samples, SOMs are also used to improve the completeness of spectroscopic sample, the representativity of which is pivotal to calibrate the photometric redshifts \citep[e.g. the C3R2 spectroscopic calibration sample for \textit{Euclid}, see][]{mastersMappingGalaxyColorRedshift2015,euclidcollaborationEuclidPreparationXX2022}.
In addition to SOMs, other unsupervised dimensionality reduction techniques have been explored to study large datasets, such as t-SNE \citep{steinhardtMethodDistinguishQuiescent2020}, locally-linear embedding \citep{vanderplasReducingDimensionalityData2009} or autoencoders \citep{portilloDimensionalityReductionSDSS2020}.

\begin{figure*}
    \centering
        \centering
        \includegraphics[width=0.49\linewidth]{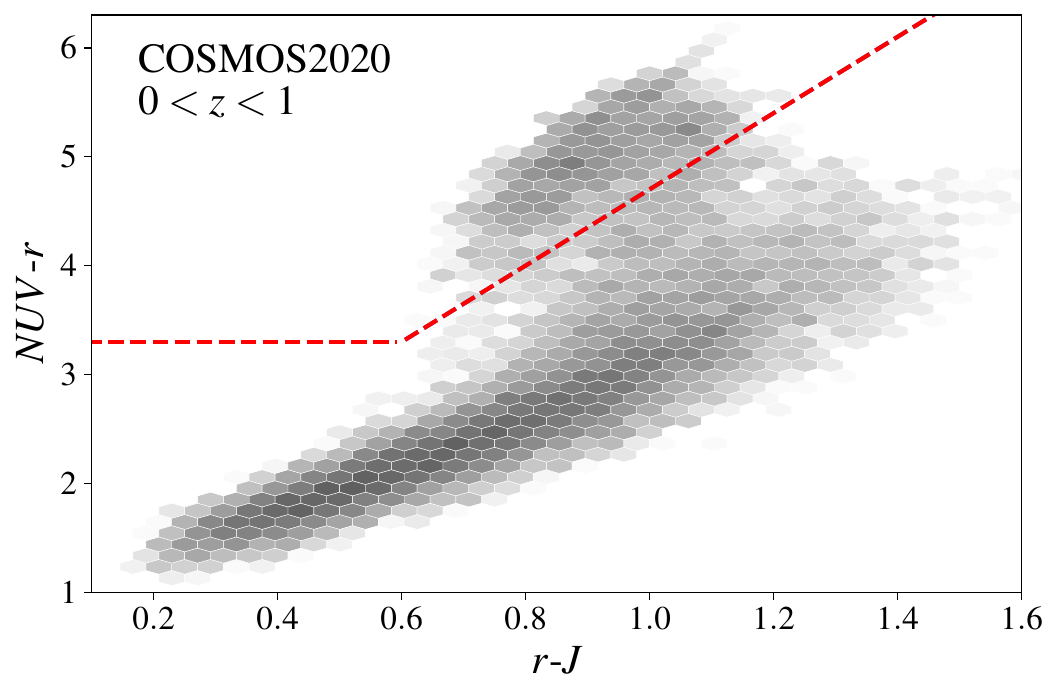}\includegraphics[width=0.48\linewidth]{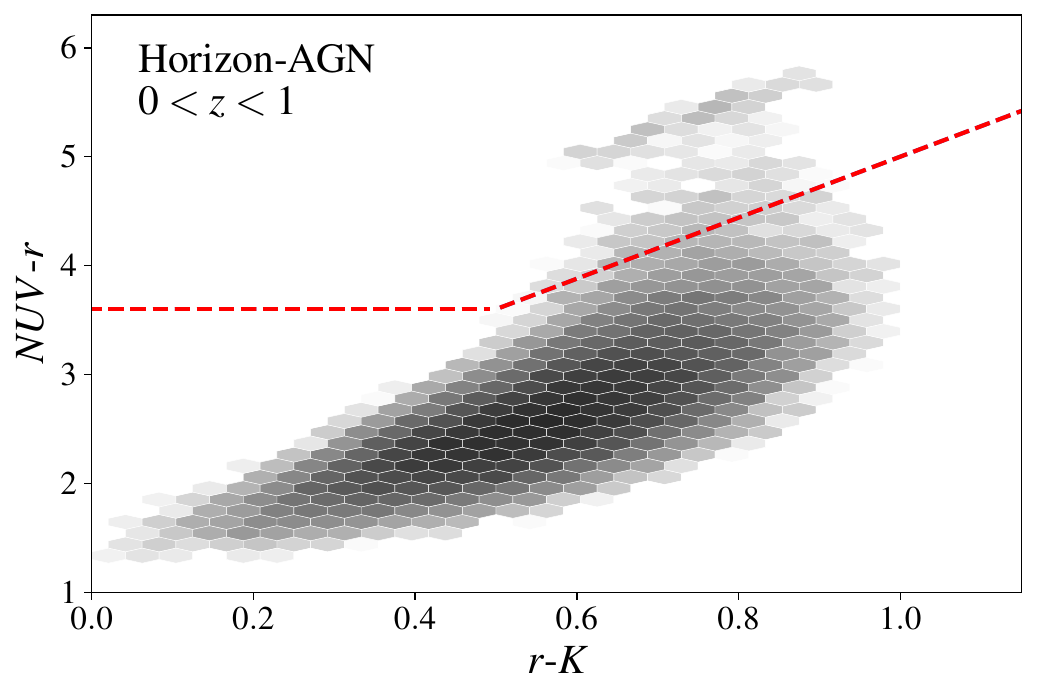}

    \caption{{Colour-colour diagram used to distinguish passive from star-forming galaxies in the COSMOS2020 (\textit{left}, ${\rm NUV}-r$ versus $r-J$ rest-frame plane) and \hagn{} (\textit{right}, ${\rm NUV}-r$ versus $r-K_{\rm s}$ rest-frame plane) catalogues. Here we display galaxies in the redshift range $0<z<1$ and with mass larger than $10^{9.5}\,\si{\Msun}$. Passive galaxies are located in the top left corner.}
    }\label{fig:nrk}
\end{figure*}
Implicitly, these works assume that the manifold occupied by galaxies in photometric magnitude space can be projected in typically two dimensions without significant loss of information.
While this implicit assumption is convenient for visualisation purposes, the choice of two dimensions isn't based on any rigorous argument.
Indeed, the question of the complexity of galaxy formation is still open. Several studies have used a principal component analysis (PCA) to quantify the number of parameters needed to capture the variance in the observed properties of galaxies, with the main findings being that only a handful are required (in observations of 21cm emission,~\cite{disneyGalaxiesAppearSimpler2008}, in SDSS spectra,~\cite{sharbafWhatDrivesVariance2023}). One limitation of using PCA is however that it is a linear method, and thus cannot capture non-linear correlations between the observables, and should thus be interpreted as an upper limit on the actual dimensionality of the data.
Similarly, machine learning approaches suggest that galaxies lie on a low-dimensionality manifold \citep{villaescusa-navarroCosmologyOneGalaxy2022,echeverriCosmologyOneGalaxy2023,chawakCosmologyMultipleGalaxies2023}.
Taken at face value, this is at odds with theory, where models of galaxy formation involve complex physical mechanisms \citep[see e.g.][]{somerville_PhysicalModelsGalaxy_2015}. Those depend, at least, on the mass of the halo, its angular momentum distribution, the impact parameter and timing of the mergers, and the non-linear physics of gas cooling, star formation, and stellar/AGN feedback processes, suggesting that galaxy formation unravels in high-dimensional parameter space.

The complexity of galaxy formation has however been recently questioned. Indeed, simulations have revealed that, at fixed galaxy mass, the scatter in properties such as disk sizes may be driven by the misalignment between the inertia tensor and tidal tensor in the initial conditions \citep{cadiouStellarAngularMomentum2022,moonMutualInformationGalaxy2024}, potentially explaining the origins of scatter in well-established  scaling relations such as the size-magnitude relation \citep[][]{Courteau07}. In this picture, galaxy formation is a self-regulated process, where only a subvolume of the parameter space is accessible by the global properties of galaxies.

\begin{figure*}
    \centering
        \centering
        \includegraphics[width=0.98\linewidth]{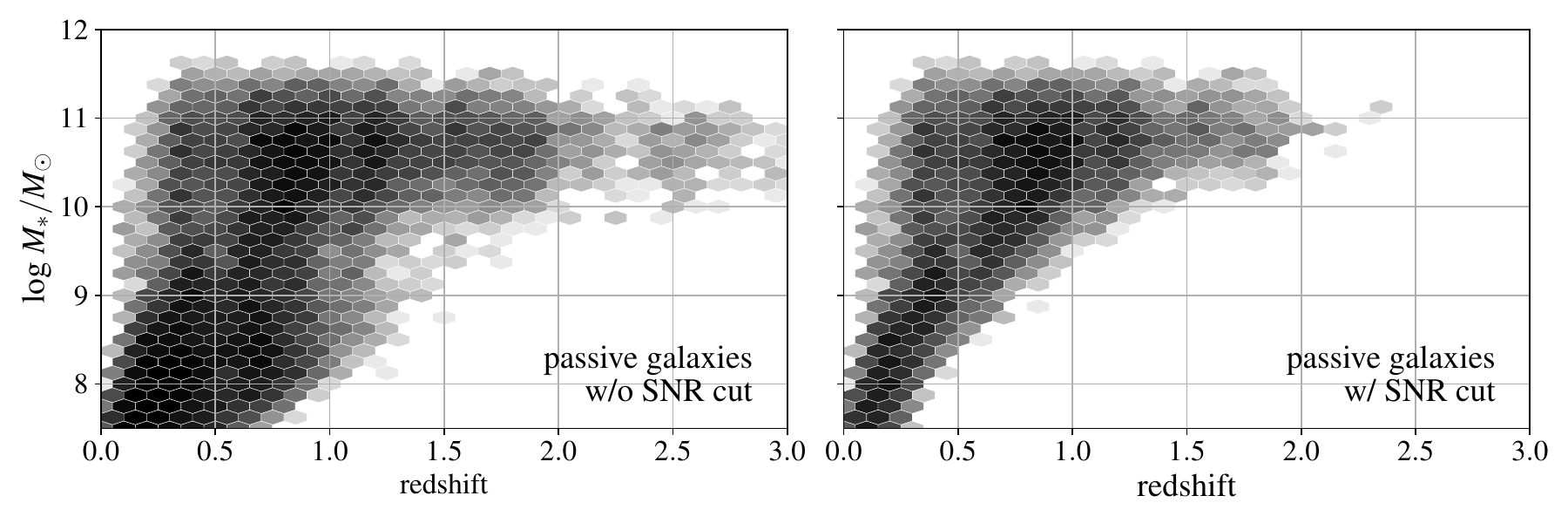}
    \caption{
        The mass and redshift distribution of passive galaxies in the whole sample (\textit{left}) and after having applied the ${\rm SNR}>2$ cut (\textit{right}).
        Our sample is biased towards high-mass and low redshift passive galaxies, for which significant flux may be detected in all bands.
    }\label{fig:passgal}
\end{figure*}
In this paper, we aim to answer the question of the dimensionality of the manifold occupied by galaxies in the multi-dimensional space of their photometric magnitudes.
We rely on observations of hundreds of thousands of galaxies from the COSMOS catalogue, and simulated ones from the \hagn simulation; we describe these datasets in \cref{sec:catalogue-data}.
We then present our parameter-free method, based on the approach of~\cite{granataAccurateEstimationIntrinsic2016}, to estimate their intrinsic dimensionality in \cref{sec:dimensionality-estimation}.
We discuss our results in \cref{sec:results} and wrap up in \cref{sec:conclusion-and-discussion}.

\section{Catalogue data}
\label{sec:catalogue-data}

\subsection{The COSMOS2020 catalogue}
\label{ssec:the-cosmos2020-catalogue}
We rely on the COSMOS2020 catalogue \citep{weaverCOSMOS2020PanchromaticView2022} to estimate the dimensionality of galaxy photometry. COSMOS2020 is a deep multi-wavelength catalogue extracted from the COSMOS field. It includes the photometry in 35~filters from Ultraviolet (UV) to infrared (IR). The improvement of this catalogue with previous versions is especially driven by deeper optical, NIR and IR data obtained with the HSC camera \citep{miyazakiHyperSuprimeCamSystem2018} on the Subaru telescope, the VIRCAM instrument on the VISTA telescope \citep[as part of the UltraVISTA survey, see][]{mccrackenUltraVISTANewUltradeep2012,euclidcollaborationEuclidPreparationXVII2022}, and the IRAC instrument on the Spitzer telescope \citep[as part of the SPLASH survey, see][]{steinhardtStarFormationSpitzer2014}. %
In the following, we use the {\sc CLASSIC} version of the catalogue, that is, the one that uses the photometry obtained with {\sc SExtractor} \citep{bertinSExtractorSoftwareSource1996}. More specifically, galaxies were identified from a $izYJHK_{\rm s}$ chi-squared image, and magnitudes are extracted in a \ang{;;2} aperture {reaching a 3$\sigma$-depth of 28.1 and 25.3 in the $g$-band and the $K$-band respectively}. A full description of the catalogue preparation can be found in~\cite{weaverCOSMOS2020PanchromaticView2022}. We remove from the sample all galaxies that have been flagged as being near bright sources and areas affected by artefacts. The total area after accounting for these masks is \SI{1.27}{deg^2}. We use galaxy properties (redshifts, masses and rest-frame $UVJ$) derived through the {\sc LePhare} code \citep{ilbert06}.

\subsection{The \hagn{} mock catalogue}
\label{ssec:the-mock-catalogue}
We work with a simulated catalogue to investigate the effect of noise in the data, and to explore how different is the dimensionality of galaxy photometry when estimated either from their apparent or absolute magnitudes.

The \hagn{} hydrodynamical simulation \citep{dubois_DancingDarkGalactic_2014}, run with the adaptive mesh refinement code RAMSES \citep{teyssierCosmologicalHydrodynamicsAdaptive2002}, was used to produce a mock catalogue with the same properties as the observed COSMOS2020 sample. \hagn{} has a box size of \SI{100}{Mpc/\hred} and a DM mass resolution of \SI{8e7}{\Msun}. In brief, the simulation includes gas heating, cooling, star formation, feedback from stellar winds, type Ia and type II supernovae with mass, energy, and metal release in the interstellar medium, and black hole formation. Depending on the accretion rate, AGNs release energy in a quasar or radio mode. A light cone was extracted on the fly to build realistic mocks\footnote{a version of which being publicly available at \href{https://www.horizon-simulation.org/data.html}{Horizon-AGN Virtual Observatory}.} \citep{laigle_HorizonAGNVirtualObservatory_2019,gouinWeakLensingHorizonAGN2019}.

Galaxies were extracted based on the position of stellar particles using the {\sc AdaptaHOP} structure finder \citep{aubert_OriginImplicationsDark_2004, tweed_BuildingMergerTrees_2009}. Mock spectra were produced for each galaxy, assuming~\cite{bruzualStellarPopulationSynthesis2003} synthetic stellar population models and a Chabrier initial mass function \citep{chabrierGalacticStellarSubstellar2003}. Dust attenuation was implemented for each star particle using the gas-metal mass distribution along the line-of-sight as a proxy for the dust distribution. All details about the spectra computation can be found in~\cite{laigle_HorizonAGNVirtualObservatory_2019}. In addition to spectra, both absolute and apparent magnitudes were extracted in the same photometric filters as available in the COSMOS2020 catalogue. Mock photometric fluxes were perturbed by assuming a combination of Poisson noise and Gaussian noise, considering in each filter the same depth as the COSMOS2020 catalogue, so that galaxies of similar luminosity have a similar signal-to-noise ratio (SNR) in the mock and in the observed catalogues.

Overall, the galaxy populations agree well with the observations, but some discrepancies must be mentioned \citep{kaviraj_HorizonAGNSimulationEvolution_2017}. \hagn{} produces galaxies that are too massive for their halo, in particular at low masses and low redshift. As a result, the stellar-to-halo mass ratio is up to 10 times larger than expected observationally in haloes of mass \SI{e11}{\Msun} \citep{shuntovCOSMOS2020CosmicEvolution2022}. Furthermore, massive passive galaxies at low redshift are not as numerous as in observations, due to residual star formation in massive galaxies, and low-mass galaxies are not star-forming enough \citep[see e.g.][]{dubois_HORIZONAGNSimulationMorphological_2016}.

\subsection{Dataset selection}
\label{ssec:dataset-selection}

\paragraph*{Filter set:} We estimate the dimensionality of the galaxy distribution based on the following fourteen fiducial broad-band filters: $u^*, g, r, i, z, y, Y, J, H, K_{\rm s}, B, V$ and IRAC's CH1 and CH2. The full definition of these filters is provided in~\cite{weaverCOSMOS2020PanchromaticView2022}. Unless stated otherwise, this set of fourteen filters is the one used throughout the analysis.
However, because of their width, photometry in broad-band filters cannot easily isolate the emission by the continuum from the nebular emission lines. The latter, which could carry additional information with respect to the continuum, can be captured when adding narrow-band filters to the filter set. To that end, we also consider the fourteen narrow-band filters are also available in the COSMOS2020 catalogue in the range $\left[ 0.5, 0.8 \right]\si{\micro m}$. We will also quantify in the following the change in dimensionality due to their addition to our fiducial filter set.

\paragraph*{Passive galaxy identification:}
{Passive galaxies in  \hagn{}  were identified from their position in the ${\rm NUV}-r$ versus $r-K_{\rm s}$ rest-frame plane. More specifically, we define passive galaxies as those with $\left({\rm NUV}-r >2.8(r-K_{\rm s})+2.2 \right) \& \left( {\rm NUV}-r > 3.6\right)$. This empirical definition provides the best segregation between passive and star-forming galaxies in the \hagn{} simulated light cone. In observations, we adopt the usual definition based on the ${\rm NUV}-r$ versus $r-J$ rest-frame plane, to match previous observational works \citep{ilbertMassAssemblyQuiescent2013,laigleCOSMOS2015CatalogExploring2016,weaver23}, with $\left({\rm NUV}-r >3(r-J)+1.2 \right) \& \left( {\rm NUV}-r > 3.1\right)$.
We show the colour-colour diagrams on \cref{fig:nrk} together with our selection for the COSMOS and \hagn{} datasets.
We also note that the definition of passive galaxies is slightly different in the simulated and observed datasets, to account for the difference outlined in \cref{ssec:the-mock-catalogue} in terms of star-formation between the observed and simulated populations}.
However, the method we propose in the following provides an ensemble average of the dimensionality and is thus insensitive to where exactly the cuts are done, as long as they are sufficiently far from the bulk of the distribution.

\begin{figure*}
    \centering
    \begin{minipage}{0.3\linewidth}
        \centering
        \includegraphics[width=\linewidth]{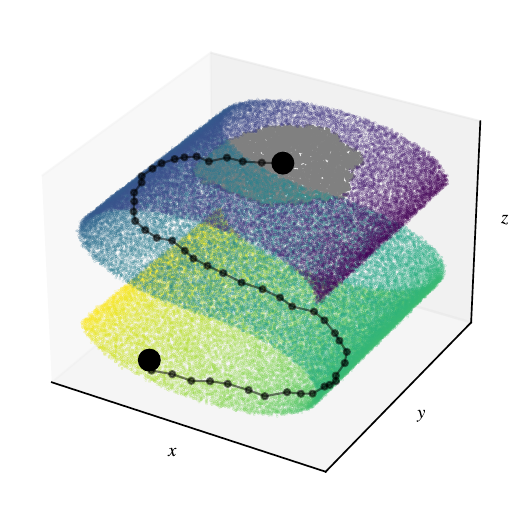}
        {\bf (a)}
    \end{minipage}\hspace{2cm}%
    \begin{minipage}{0.5\linewidth}
        \centering
        \includegraphics[width=\linewidth]{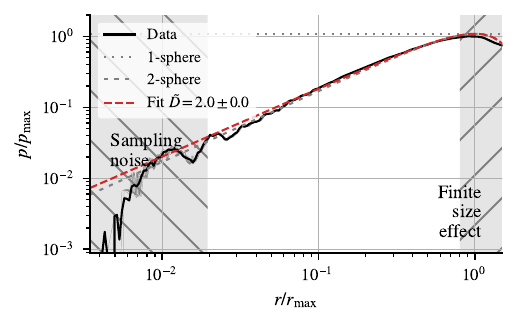}
        {\bf (b)}
    \end{minipage}
    \caption{
        The dimensionality of a dataset can be estimated from the distribution of distances between points.
        On the left {\bf (a)} the distribution of \num{100 000} points drawn from a 2D S-curve and embedded in 3D.
        The distance between two points (large black dots) can be computed by finding the shortest path between them (black dotted line, we represent one point every ten for clarity) from nearest neighbour to nearest neighbour.
        We show all the points within a distance $r$ from the starting point in gray.
        The number of such points grows here as $r^2$; in general, it grows as $r^D$.
        By measuring the probability $p$ of finding a point within a distance $r$,
        we can thus estimate the intrinsic dimensionality of the dataset, as is shown on the right {\bf (b)}.
        We detail the procedure in the text.
    }\label{fig:scurve}\label{fig:scurve-ID}
\end{figure*}

\paragraph*{Signal-to-noise ratio:}
In order to limit the impact of noise on our results, we keep only galaxies for which ${\rm SNR}>2$ in each band in each catalogue. {We note that this cut will introduce a bias in the sample, since it imposes that galaxies emit flux even in the bluest bands. It therefore excludes those very passive and faint galaxies which remain undetected in the bluest bands, and limits the maximum redshift that we can probe. As an illustration, we show in \cref{fig:passgal} the distribution of passive galaxies as a function of mass and redshift with and without the SNR cut. Passive galaxies at high redshift and in the relatively low mass range are not present with the SNR cut. Similarly, very dusty star-forming galaxies are likely to be excluded from the sample with this SNR cut.}

We list in \cref{tab:measured_dims} all the datasets together with the selection operated, along with the number of galaxies in each sample.
We note that, for the sake of consistency, we use the same subset of simulated galaxies that have a ${\rm SNR}>2$, including in the datasets with no simulated noise. We checked that including galaxies with ${\rm SNR}<2$ does not change the results obtained for the mock dataset with no simulated noise, and yields consistent results for the COSMOS catalogues, albeit with larger error bars.
We however note that those faint galaxies may have an intrinsically different dimensionality. We however find that, should that be case, the signal would be hidden beyond the relatively high noise level.

\section{Dimensionality estimation}
\label{sec:dimensionality-estimation}

In this work, we aim to measure the intrinsic dimensionality $D$ of wide-band photometric data of galaxies.
Our data is comprised of $N$ bands with a sample size of $N_\mathrm{g}$ galaxies.
    To that end, we rely on intrinsic dimensionality estimation methods.
    We note that, although intrinsic dimensionality estimation bears similarities with dimensionality reduction techniques (Variational Autoencoders, PCA, t-SNE, etc.), their goal differ fundamentally: while dimensionality reduction techniques aim at finding a low-dimensional representation of the data that preserves the topology of the data, intrinsic dimensionality estimation aims at finding the number of independent parameters that are needed to describe the data. The latter can then be used to inform the former, by providing a lower bound on the dimensionality of the latent space.

    Intrinsic dimensionality estimation is a well-studied problem in machine learning and statistics, and many methods have been proposed to estimate it
    \citep[see e.g.][]{pestovIntrinsicDimensionDataset2007,gomtsyanGeometryAwareMaximumLikelihood2019,cloningerDeepNetworkConstruction2021,dadapy,glielmoDADApyDistancebasedAnalysis2022}.
    However, most of the proposed algorithms are local: the dimensionality is estimated from the statistics of points found within some neighbourhood (for example, using the distribution of the second-nearest neighbour, \citealt{glielmoDADApyDistancebasedAnalysis2022}) before being averaged out over the whole dataset. As such, they are not optimal for astronomical datasets where the noise in the data is large, and hence local estimations equally noisy.
    Alternatively,~\cite{granataAccurateEstimationIntrinsic2016} proposed an algorithm that allows to estimate the global -- rather than local -- intrinsic dimensionality of a dataset that is robust to local sampling noise and noise in the data (we will discuss this later).
    In addition to its robustness, this method has the advantage of being rather simple and of having a geometrical interpretation which, we hope, will prove useful to the reader. We delay the application of more involved methods \citep[e.g.,][]{stubbemannIntrinsicDimensionLargeScale2023} to future work.

Let us present this method qualitatively.
If the distance between points in the feature space is given by the function $d$, let us define the probability of finding a point at separation $r$ as
\begin{equation}
    p(r) = \frac{2}{N(N-1)} \sum_{i<j} \delta_\mathrm{D}(d(\vec{x}_i, \vec{x}_j)-r),
    \label{eq:pr}
\end{equation}
where $\delta_\mathrm{D}$ is the Dirac delta function and $\vec{x}_i$ is the data-vector of the $i{}^{\mathrm{th}}$ data point.
Note that, here, one could approximate the distribution another kernel (e.g., a Gaussian kernel) instead of a Dirac delta function. While this would allow to accommodate for noise, we found it to have a negligible impact on the results\footnote{As will be shown on \cref{fig:mock_vs_COSMOS}, changing the kernel would affect the left side of $p(r)$ which we discard anyway.}.
If the manifold is compact, as is the case for photometric data, this probability eventually falls to zero as there are no points infinitely far from one another.
Similarly, the probability of finding two points vanishes at small separations, unless the intrinsic dimension is null as well.
Indeed, at small separations, the number of points within a distance $r$ from a point at position $\vec{x}$ goes as $r^D$, so $p(r)$ scales as $r^{D-1}$.
If the distance function $d$ is the Euclidean distance in the embedding space,~\cite{eckmannFundamentalLimitationsEstimating1992} showed that the number of points required for accurate measurement of the intrinsic dimensionality scales as $N \sim 10^{D/2}$.
This is a direct consequence of the well-known curse of dimensionality and the fact that limiting the analysis to the low-separation limit of $p(r)$ discards a large fraction of the data.

The procedure can however be improved by addressing these two issues, as was shown in~\cite{granataAccurateEstimationIntrinsic2016}.
First, if the manifold is folded onto itself, as illustrated \cref{fig:scurve}, $p(r)$ will display peaks at separations corresponding to the separation between the folds.
This can be addressed by substituting the Euclidean distance between points with a graph distance, which is the length of the shortest path that connects them through the graph of $k$-nearest neighbours; we illustrate it in \cref{fig:scurve}.
This essentially allows to `unroll' the manifold and regularises $p(r)$, which~\cite{granataAccurateEstimationIntrinsic2016} showed to then be well approximated by the distribution of a $D$-dimensional hyper-sphere up to the maximum of $p(r)$.
The latter has a simple closed-form expression, $p_{\mathrm{sphere},D}(r) \propto \sin^{D-1}(r\pi/2)$, which we can fit to the data.
This estimator was shown to yield robust estimates of the dimensionality of a 20-dimensional dataset from a few thousand points.

In our case, we seek to measure the intrinsic dimensionality of data with shape $(N_\mathrm{g}, N)$, where $N_\mathrm{g}$ is of the order of \num{10 000} to \num{100 000} and $N=14$ wide photometric bands, which is well within the reach of the method.
To that end, we first rescale the magnitudes into the range $[0, 1]$ using the same mapping for all bands, and we build a KD-Tree using the Euclidean distance in the embedded space.
We then build the connectivity graph, where edges connect each data point to its $k$-nearest neighbours, increasing $k$ from one until the graph is fully connected.
Ideally, one would then compute the graph distance between all pairs of points (for example with the Floyd-Warshall algorithm, with complexity $\mathcal{O}(N_{\rm g}^3)$). This however proved to be prohibitively expensive and we instead resort to using a small subset of points. To that end, we pick $N_\mathrm{sub} = 200$ points at random\footnote{We found 200 points to be sufficient to have a converged estimate of the dimensionality while remaining computationally affordable.} and compute the graph distance from each of those to all others using Dijkstra's algorithm (with complexity $\mathcal{O}(N_\mathrm{sub} \times N_{\rm g}\log N_{\rm g})$, \citealt{dijkstraNoteTwoProblems1959}). Plugging these distances back into \cref{eq:pr} yield estimates of the pairwise distance probability $p(r)$.
We split this sample of 200 estimates of $p(r)$ into ten subsamples, and we estimate the intrinsic dimensionality of each subsample as follows.
For each of those, we normalise $p(r)$ by its maximum value and $r$ by the corresponding radius.
We then perform a least-square regression with the analytical expression of a $D$-dimensional hyper-sphere, $p_{\mathrm{sphere},D}(r/r_\mathrm{max}) = {C}\sin^{D-1}(r\pi/2 r_\mathrm{max})$, with free parameters ${D}$ and ${C}$. On the low-separation end, we remove radii that contain less than a thousand pairs to limit shot-noise. On the high-separation end, we truncate at $r=0.8 r_\mathrm{max}$ where the fit to the hyper-sphere started diverging significantly.
We note that our results are robust to changes to these choices.
We obtain our estimated dimensionality and its uncertainty by taking the mean and standard deviation of the ten estimates.

\begin{figure}
    \centering
    \includegraphics{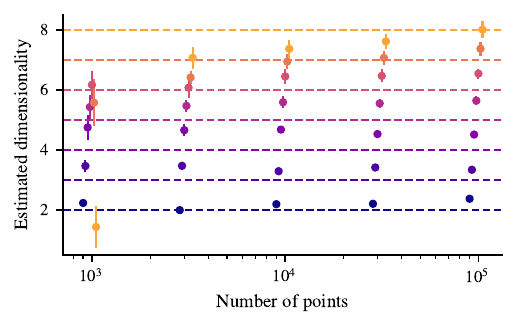}
    \caption{
       We use mock datasets that consist of points drawn from $D$-dimensional balls and measure their dimensionality as a function of the number of sample points.
       The ground truth values are shown as dashed lines with our estimates of the dimensionality as points with matching colour; the error bars represent the 1-$\sigma$ dispersion.
       The accuracy of the dimensionality estimation increases with the increasing number of points.
       For this particular sample, our method is biased towards overestimating the actual dimensionality, but it is able to correctly distinguish datasets with an intrinsic dimensionality of up to eight.
    }\label{fig:ball_vs_Npt}
\end{figure}

As an illustrative example, we show on \cref{fig:scurve-ID}, right panel, our estimate of $p(r)$ for the S-curve dataset presented in \cref{fig:scurve}, left panel, sampled here with \num{100 000} points and embedded in 3D. We recover here an excellent agreement between the measured dimensionality and the truth value, 2.
To assess the accuracy of our method, we then draw \num{100 000} points from a $D$-dimensional ball embedded in a $N$-dimensional space. We chose here $N=14$ and vary $D$ from $1$ to $8$. In practice, we uniformly draw points within a $D$-dimensional unit ball and fill the remaining dimensions with a small ($10^{-2}$) uniformly distributed value to mimic the effect of a small signal-to-noise.
\Cref{fig:ball_vs_Npt} shows the estimated dimensionality together with the truth value as a dashed line.
We see that the accuracy of the measurement depends on the intrinsic dimensionality, with the most accurate measurements obtained for low-dimensional datasets, and on the number of points used to estimate it.
We also note that our estimator is biased towards an overestimation of the actual intrinsic dimensionality by up to $0.5$.
While the bias may depend on the exact geometry of the manifold, we however note that, for a fixed geometry, our estimator can correctly distinguish datasets with different dimensionality.
In practical terms, with a sample size of \num{100 000} points, we can distinguish datasets with up to seven intrinsic dimensions within one standard deviation.
With a sample size of \num{10 000}, we can distinguish datasets with up to six intrinsic dimensions.

In the following, we will mostly focus on the variation of the measured dimensionality, so that we limit the extent to which the bias is the estimator contaminates our conclusions.

\begin{table*}
    \centering
    \caption{
        Measure of intrinsic dimensionality (ID). Here we only use high signal-to-noise ratio data.
        Contributions to ID: (1) measurement noise, (2) stellar mass, (3) redshift.
        For star-forming (SF) galaxies, an extra parameter is (4) the sSFR.
    }\label{tab:measured_dims}
    \begin{threeparttable}
        \begin{tabular}{lllS[table-format=6.0]Sr}
            \bf Dataset & \bf Comment & \bf Selection & {\bf Size} & {\bf Dimensionality} & \bf Figure \\
            \toprule
            HAGN            & & -            & 715006 & \num{4.3(0.2)} & \ref{fig:mock_vs_COSMOS}\\
                            & & star-forming & 706720 & \num{4.5(0.3)} & \\
                            & & passive      &   8286 & \num{2.9(0.2)} & \\[.2em]
            &no noise\tnote{*}& -                                                        & 715006 & \num{3.3(0.1)} & \\
                           &no noise\tnote{*} & star-forming                                             & 706720 & \num{3.3(0.1)} & \ref{fig:HAGN-passive_vs_SF}\\
                           &no noise\tnote{*} & passive                                                  &   8286 & \num{2.2(0.3)} & \ref{fig:HAGN-passive_vs_SF}\\
                           &no noise\tnote{*} & $10.0\leq \log_{10}(M_\star/\mathrm{M_\odot}) \leq 10.5$          & 105296 & \num{2.3(0.1)} & \\
                           &no noise\tnote{*} & $10.5\leq \log_{10}(M_\star/\mathrm{M_\odot}) \leq 11.0$          &  28300 & \num{2.4(0.1)} & \\
                           &no noise, absolute magnitudes\tnote{*} & -                                   & 715006 & \num{2.2(0.1)} &  \\
                           \midrule
            COSMOS          &&  -                                                             & 393118 & \num{4.3(0.2)} & \ref{fig:mock_vs_COSMOS}\\
                            && star-forming                                                   & 359159 & \num{4.6(0.2)} & \ref{fig:COSMOS-SF-vs-passive}\\
                            && passive                                                  & 30260 & \num{2.8(0.1)} & \ref{fig:COSMOS-SF-vs-passive}\\[.2em]
                            && $10.0\leq \log_{10}(M_\star/\mathrm{M_\odot}) \leq 10.5$ & 32190 & \num{3.0(0.2)} &  \\
                            && $10.5\leq \log_{10}(M_\star/\mathrm{M_\odot}) \leq 11.0$ & 15581 & \num{3.2(0.2)} &  \\
        \end{tabular}
        \begin{tablenotes}
            \item[*] For consistency, we select the simulated galaxies that have a large signal-to-noise in the catalogue with simulated noise.
        \end{tablenotes}
    \end{threeparttable}
\end{table*}

\section{Results}
\label{sec:results}

\begin{figure}
    \centering
    \textbf{Wide-band COSMOS data}
    \includegraphics[width=\linewidth]{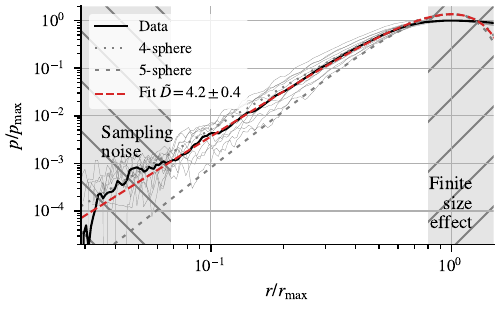}
    \textbf{Wide-band Mock data}
    \includegraphics[width=\linewidth]{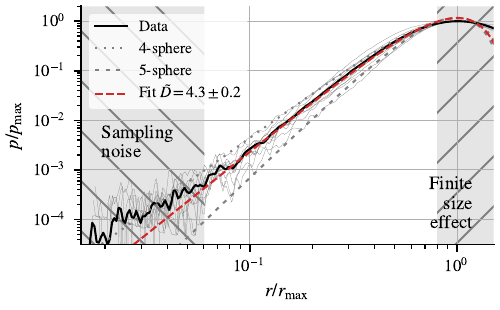}
    \caption{
        Wide-band photometric data from the COSMOS catalogue have a low dimensionality and are intrinsically four-dimensional.
        We measure the pairwise distance probability $p$ as a function of separation $r$ and compare it to the analytical formula of a $D$-dimensional hypersphere.
        We show in dashed red the result of the best fit.
        Both simulated and observed data have an intrinsic dimensionality between four and five.
    }\label{fig:mock_vs_COSMOS}
\end{figure}

First, let us compare the observed and simulated catalogues.
We measure their dimensionality and present the results on \cref{fig:mock_vs_COSMOS}, with data including all galaxies from COSMOS (top) and the \hagn{} mock catalogue (bottom). In both cases, we find that the dimensionality is around $4.2$.
Despite data from the simulation missing passive galaxies at the high-mass end and star-forming ones at the low-mass end, we, however, find that the simulated and the observed catalogues have overall similar dimensionality.
To confirm that this dimensionality is driven by actual correlations in the dataset rather than an effect of the limited size of our sample, we randomly reshuffle the entries of the COSMOS datasets independently for each column to erase possible correlations. Doing so, we find a much larger dimensionality of \num{9.2(1.2)}.
We however recall that such, for such high dimensionlities, our method starts saturating and this value should thus be interpreted as a lower bound on the actual intrinsic dimensionality of the data.
However, this confirms that the measured dimensionality in the fiducial, unshuffled, dataset reflects internal structures rather than be an artefact of the limited size of our sample.

With this in mind, we then use the noise-free magnitudes to estimate the impact noise has on the apparent complexity of the observations.
Our finding is that noise masquerades an intrinsic dimension of the dataset, so removing noise decreases the overall dimensionality by one (to $3.2\pm0.1$), and this despite having limited ourselves to galaxies with a signal-to-noise ratio larger than 2.
The simulated dataset also allows us to employ absolute, rather than apparent, magnitudes. This removes the effect of redshift from the data.
Working with noise-free apparent magnitudes, we observe that the dimensionality further decreases by one (to $2.2\pm0.1$).
We interpret this as the (unsurprising) fact that one key variable driving the magnitudes in different bands is the redshift of the galaxy.

To gain further insight into the dataset, let us now split the galaxies between star-forming and passive ones, as described in \cref{ssec:dataset-selection}. Doing so, we find that passive galaxies have a dimensionality of $2.8\pm0.1$ ($2.9\pm0.2$ in the simulated dataset), while star-forming galaxies have $4.6\pm0.2$ ($4.5\pm0.3$ in the simulated dataset).

We then include in our datasets the 14 narrow-band filters, with the expectation those would encode independent information about the star-formation activity of the galaxies.
We however find that the addition of those filters does not change significantly the estimate of the intrinsic dimensionality which goes from \num{4.2(0.4)} to \num{4.5(0.3)} when including the narrow-band filters. We reach a similar conclusion for star-forming galaxies (from \num{4.6(0.2)} to \num{4.6(0.3)}) or passive ones (from \num{2.8(0.1)} to \num{2.8(0.2)}).

Finally, we also apply our estimator to spectra in the mock catalogue, finding a dimensionality of \num{5.0(0.3)}.
We warn that a careful analysis of the performance of our method should be carried out
in the limit of large embedding dimensions (here, 2000) to be able to reach definitive conclusions.
Notwithstanding, the measured dimensionality here is larger than the corresponding dimensionality of the wide-band photometric dataset (which is \num{3.2(0.1)}).
We interpret this as evidence that spectra are more information-rich and that at least two additional parameters may be extracted from them.
We delay further exploration of spectra to the future.

\section{Conclusion and discussion}
\label{sec:conclusion-and-discussion}

The goal of this paper was to estimate the dimensionality of broad-band photometric data.
To that end, we have adapted the dimensionality-measurement technique of~\cite{granataAccurateEstimationIntrinsic2016}.
Our main results are the following:
\begin{enumerate}
    \item The intrinsic {large-scale} dimensionality of {observed} broad-band data is between four and five. Mocks are consistent with observations {when similar level of noise and cut are applied}.
    \item Realistic estimates of the noise in the mock catalogue account for one of these dimensions.
    \item Star-forming galaxies are intrinsically more complex ($D\sim 3-4$) than passive galaxies ($D\sim 2-3$).
\end{enumerate}

Our results highlight that most of the variance in global, large-scale properties of galaxies can be described with a small number of independent parameters.
This is in line with similar findings of~\cite{villaescusa-navarroCosmologyOneGalaxy2022,chawakCosmologyMultipleGalaxies2023}.
These previous studies relied on the use of machine-learning and focused on physical parameters (masses, radii, velocities) rather than observational ones.
Our method aims at providing the number of `main' parameters driving scaling relations, which we showed here to be consistent with mass, redshift, and star formation rate.
In the future, we suggest extending the mock dataset to mix observational information with physical quantities, such as mass, radius, or spin.
Any increase in the measured dimensionality would then suggest that the physical quantities cannot be derived from the observational ones alone.
We delay such analysis to future work, as this would require carefully rescaling physical and observational parameters to allow building a similarity metric between them.

We emphasize that relatively small deviations from scaling relations are only revealed once the driving effect of these main parameters is removed.
For example, when we bin by stellar mass in the noise-free absolute-magnitude simulated dataset, we find that this does not result in a decrease in the number of dimensions.
Indeed, once the main parameters driving the absolute magnitudes were fixed, we started probing secondary parameters that drive the scatter around scaling relations, such that the dimensionality measurement is now probing smaller variations in magnitudes.

Our approach stands out compared to more traditional PCA-based approaches \citep[e.g.][]{2004AJ....128.2603Y,2006MNRAS.370..828F,portilloDimensionalityReductionSDSS2020,sharbafWhatDrivesVariance2023} or singular-value decomposition \citep[SVD,][]{coorayCharacterizingUnderstandingGalaxies2023}.
For example, \cite{sharbafWhatDrivesVariance2023} employed a principal component analysis (PCA) to show that most of the variance in SDSS spectra can be explained by a small number of components ($\lessapprox 3$), yielding a similar result as ours. PCA-based approaches have the advantage of being relatively simple and efficient, yet they suffer from two main limitations. First, the analysis on the number of components depends on the fraction of the total variance one wants to reproduce. Second, PCAs cannot reproduce non-linear relations between variables, as we showed on \cref{fig:scurve} (PCA would yield three components, while the true dimensionality is two).
In comparison, our non-parametric approach is able to capture non-linear relations between variables and therefore relaxes the assumption that the data is linearly separable and well-represented by a multivariate Gaussian distribution.

\cite{ferrerasEntropyGalaxySpectra2023b} uses information theory to recover information about the stellar component. One of the authors' main findings is that the information content of spectra is rather low, with most of the information being contained in a few components (\SI{4000}{\Angstrom}-break and Balmer absorptions lines for stellar components).
In our paper, we focus on measuring the intrinsic dimensionality of the data, which has strong connection with information content: larger numbers of intrinsic dimensions require more information to fully encode the data.
Our findings are consistent with theirs in that the intrinsic dimensionality of the data is low, although a direct comparison is not possible due to the difference in the input data (wide-band photometry vs.\ spectra) and targeted inferred properties (global galaxy properties vs.\ stellar properties).

The exact dimensionality revealed here provides guiding numbers to employ dimensionality-reducing techniques, such as self-organised maps \citep{davidzonHORIZONAGNVirtualObservatory2019,davidzonCOSMOS2020ManifoldLearning2022,teimooriniaMappingDiversityGalaxy2022,sanjaripourApplicationManifoldLearning2024} locally-linear embedding \citep{vanderplasReducingDimensionalityData2009} or variational auto-encoders \citep{portilloDimensionalityReductionSDSS2020,pirasRepresentationLearningApproach2024}. Indeed, the latent space should have at least four to five dimensions to ensure the projection doesn't lose information.

Our results support an emerging scenario in which the main global galaxy properties evolve towards a low-dimensional attractor.
This was first suggested by~\cite{disneyGalaxiesAppearSimpler2008}, although our findings support more complexity than the one-dimensional attractor they reported. We however note that our datasets rely on photometry -- and thus are affected by redshift -- while theirs did not. In addition, they excluded colour information, and thus information about star-formation rate, from their analysis, while we found this information to be relevant in driving the dimensionality up.
The closest comparison is the work of \cite{coorayCharacterizingUnderstandingGalaxies2023} that utilize data from GALEX, SDSS and UKIDSS in 11 wide-band photometric bands. Using SVD, they find that more than \SI{90}{\percent} (but not all) of the variance in the data can be explained with only two components. Although we concur in that the data can be well-represented by a low-dimensional manifold, we emphasize that our reported intrinsic dimensionality of 3-4 cannot be compared directly to their 2 components.

This idea that galaxies are governed by a small number of parameters has found renewed interest with recent research suggesting that one previously unknown key parameter driving galaxy properties is the misalignment between the inertia tensor and the tidal tensor in the initial conditions, $\tau$.
This parameter correlates well with properties in the evolved Universe such as spin and formation time \citep{moonMutualInformationGalaxy2024} or the morphology and angular momentum support of galaxies at high-redshift \citep{parkFormationMorphologyFirst2022}.
In~\cite{cadiouStellarAngularMomentum2022}, it was additionally shown to be a cause-and-effect relationship, with higher values of the misalignment yielding less bulgy and more disky, more extended galaxies at fixed halo and stellar masses.
This provides four independent parameters -- (halo) mass, local density, gravitational shear \citep[$q$, see e.g.][]{paranjape_HaloAssemblyBias_2018} and tidal misalignment ($\tau$) -- that have been shown to correlate with galaxy properties.
Remarkably, this number of independent parameters is consistent with the number of dimensions we obtained from photometric surveys.

This raises the wider question of whether this set of four parameters is enough to build a complete theory capable of fully predicting the wide-band photometric survey.
While this goes beyond the scope of this paper, we note that large cosmological volume simulations are finally ripe to tackle this problem from a statistical standpoint, as we did in this paper.
In parallel, genetic modifications of the initial conditions \citep{rey_QuadraticGeneticModifications_2018,stopyraGenetICNewInitial2021,cadiouAngularMomentumEvolution2021,cadiouCausalEffectEnvironment2021,Rey2023}, offer avenues to confirm the causal relation between those parameters and observables.

\subsection{Comparison to other works}

Compared to previous work \citep{teimooriniaMappingDiversityGalaxy2022,davidzonCOSMOS2020ManifoldLearning2022,sanjaripourApplicationManifoldLearning2024} that utilize self-organized maps (SOMs) with the latent-space dimension as an input parameter, we provide here a measurement of what this latent space should be.
\cite{zeraatgariMachineLearningbasedPhotometric2024} compared different machine learning-based approaches to classify galaxies, quasars, emission-line galaxies and stars from photometric data. In particular, they use $k$-nearest neighbours (KNN) to classify galaxies that are nearby in feature space as belonging to the same class (see also \citealt{ascasibarGalaxiesFormSpectroscopic2011} for a similar work using spectra). Our approach relies on a generalization of this, where the we replace the Euclidean distance with an approximation of the geodesic distance on the manifold, similar to the UMAP approach \citep{mcinnesUMAPUniformManifold2020}.

One limitation of our proposed approach is to rely on wide-band integrated photometry rather than resolved observations or spectra. As such, it does not include spatial information (yet) as was done with convolutional neural networks \citep{buckPredictingResolvedGalaxy2021,liew-cainConstrainingStellarPopulation2021}. This prevents us from probing the intrinsic dimensionality of resolved galaxies. The main limitation is that

\section*{Acknowledgements}
\label{sec:acknowledgements}
C.C. and O.A. acknowledge support from the Knut and Alice Wallenberg Foundation, the Swedish Research Council (grant 2019-04659), and the Swedish National Space Agency (SNSA Dnr 2023-00164). C.L acknowledges  funding from the PNCG and the French Agence Nationale de la Recherche for the project iMAGE (grant ANR-22-CE31-0007). 
This work has made use of the CANDIDE computer system at the IAP supported by grants from the PNCG, CNES, and the DIM-ACAV, as well as the Infinity Cluster, both hosted by Institut d'Astrophysique de Paris. We thank Stéphane Rouberol for running them smoothly for us. 
The authors thank C.~Pichon, M.~Rey, A.~Glielmo and O.~Ilbert for fruitful discussions.
This research was supported in part by grant NSF PHY-1748958 to the Kavli Institute for Theoretical Physics (KITP).

This project has used
\textsc{gnu parallel} \citep{tange_ole_2018_1146014},
\textsc{jupyter} notebooks \citep{kluyver_JupyterNotebooksPublishing_2016},
\textsc{matplotlib} \citep{hunterMatplotlib2DGraphics2007},
\textsc{networkx} \citep{SciPyProceedings_11},
\textsc{numpy} \citep{harris_ArrayProgrammingNumPy_2020},
\textsc{scipy} \citep{2020SciPy-NMeth} and
\textsc{Astropy}\footnote{http://www.astropy.org} a community-developed core Python package and an ecosystem of tools and resources for astronomy \citep{2013A&A...558A..33A,2018AJ....156..123A,2022ApJ...935..167A}.

\section*{Data availability}
\label{sec:data-availability}
The data underlying this article will be shared on reasonable request to the corresponding author.

\bibliographystyle{mnras}
\bibliography{authors,oa,cl} %

\appendix

\section{Different cuts}
\label{sec:different-cuts}
In this section, we provide the plots corresponding to the data in \cref{tab:measured_dims}.

\begin{figure}
    \centering
    \textbf{COSMOS, Passive only}
    \includegraphics[width=\linewidth]{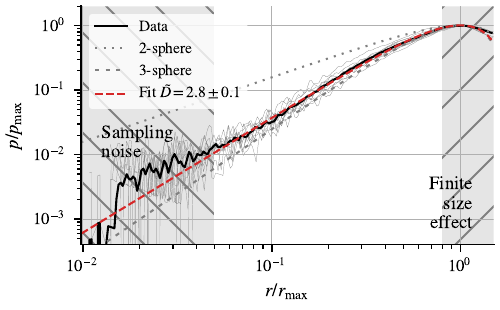}
    \textbf{COSMOS, Star-forming only}
    \includegraphics[width=\linewidth]{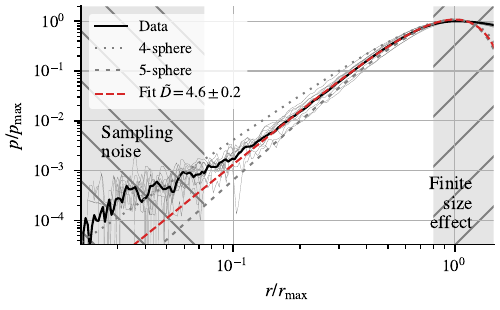}
    \caption{
        Same as \cref{fig:mock_vs_COSMOS}, but comparing the COSMOS data with only passive (top) and only star-forming (bottom) galaxies.
        Star-forming galaxies have a higher intrinsic dimensionality than passive galaxies.
    }\label{fig:COSMOS-SF-vs-passive}
\end{figure}

\begin{figure}
    \centering
    \textbf{COSMOS, Star-forming $10^{10}\leq M_\star/\mathrm{M_\odot} \leq 10^{10.5}$}
    \includegraphics[width=\linewidth]{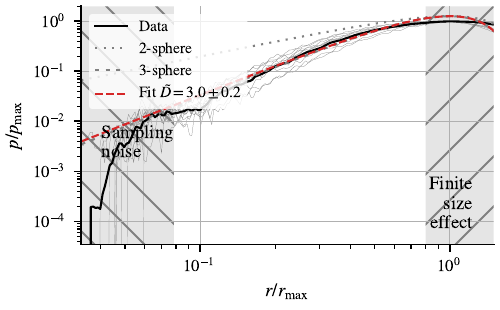}
    \textbf{COSMOS, Star-forming $10^{10.5}\leq M_\star/\mathrm{M_\odot} \leq 10^{11}$}
    \includegraphics[width=\linewidth]{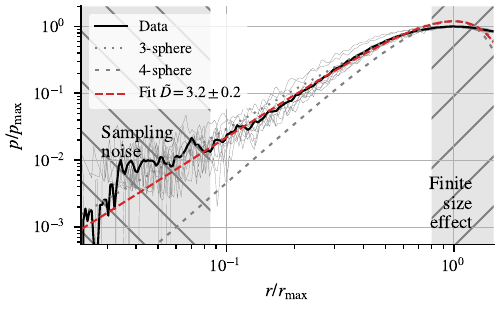}
    \caption{
        Same as \cref{fig:mock_vs_COSMOS}, but comparing the COSMOS data with only star-forming galaxies in two mass bins.
        Selecting galaxies within a narrow mass range decreases the intrinsic dimensionality by one.
    }\label{fig:COSMOS_mass_bins}
\end{figure}

\begin{figure}
    \centering
    \textbf{Wide-band mock data, passive only}
    \includegraphics[width=\linewidth]{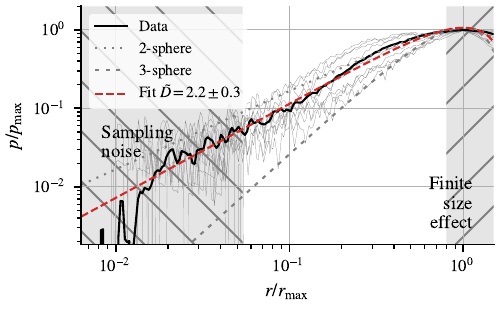}
    \textbf{Wide-band mock data, star-forming only}
    \includegraphics[width=\linewidth]{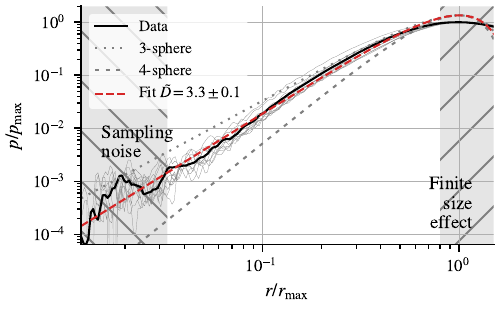}
    \caption{
        Same as \cref{fig:mock_vs_COSMOS}, but comparing the mock data with only passive (top) and only star-forming (bottom) galaxies.
        Star-forming galaxies have a higher intrinsic dimensionality than passive galaxies.
    }\label{fig:HAGN-passive_vs_SF}
\end{figure}

We show on \cref{fig:HAGN-passive_vs_SF} the dimensionality of the mock datasets including only passive galaxies (top) and only star-forming galaxies (bottom). We find that star-forming galaxies cover a region of the parameter space with higher dimensionality.

\bsp	%
\label{lastpage}
\end{document}